# The uniqueness of the integration factor associated with the exchanged heat in thermodynamics

Yu-Han Ma[1,2], Hui Dong[2], Hai-Tao Quan[3] and Chang-Pu Sun[1,2]*

**Address:** Beijing Computational Science Research Center, Beijing 100193, China

**Address:** Graduate School of China Academy of Engineering Physics, No. 10 Xibeiwang East Road, Haidian District, Beijing, 100193, China

**Address:** School of Physics, Peking University, Beijing, 100871, China

\* Correspondence: suncp@gscaep.ac.cn

**Abstract**

State functions play important roles in thermodynamics. Different from the process function, such as the exchanged heat $\delta Q$ and the applied work $\delta W$, the change of the state function can be expressed as an exact differential. We prove here that, for a generic thermodynamic system, only the inverse of the temperature, namely $1/T$, can serve as the integration factor for the exchanged heat $\delta Q$. The uniqueness of the integration factor invalidates any attempt to define other state functions associated with the exchanged heat, and in turn, reveals the incorrectness of defining the entransy $E_{vh} = C_V T^2/2$ as a state function by treating $T$ as an integration factor. We further show the errors in the derivation of entransy by treating the heat capacity $C_V$ as a temperature-independent constant.

**I. Introduction**

State functions, e.g., the internal energy and the entropy, in thermodynamics characterize important features of the system in a thermal equilibrium state [1]. Physically, some quantities are process-dependent and thus cannot be treated as state functions, e.g., the exchanged heat $\delta Q$ and the applied work $\delta W$. An integration factor $f$ can be utilized to convert the process function, e.g., the exchanged heat, into an exact differential (the change of a state function). Mathematically, the requirement of the state function can be expressed as follows: the change of the state function remains unchanged with any topological variation of the integration path on the parameter space [2]. In addition, the number of the integration factors is usually limited for any system with more than two thermodynamic variables. Such number of the integration factor is further reduced in order to define a universal state function without the dependence on the system characteristics [2]. It is a common sense that the inverse of the temperature, i.e., $1/T$, serves as the integration factor for the exchanged heat $\delta Q$, and thus a state function, the entropy can be defined. A relevant question arises here: is there any other universal integration factor associated with the exchanged heat $\delta Q$ for an arbitrary system? Several attempts on this issue have been made for both specific systems [3] and generic systems [4]. However, the uniqueness of the integration factor of interest for a generic thermodynamic system remains unexplored. In our current paper, from the first principle, we prove that only one factor, namely $1/T$, can serve as the integration factor to convert the exchanged heat into the state function. Such uniqueness invalidates any attempt to find new state functions associated with the exchanged heat $\delta Q$.

It is worth mentioning that the introduction of the so-called "state function", the entransy defined via $\delta E_{vh} = T\delta Q$ [5], in the realm of the heat transfer is an example of such attempt. The entransy is claimed to be a "state function" of a system characterizing its potential of heat transfer [5]. Since its appearance, it has triggered a lot of debates [6-10] over its validity as state function [6] as well as its usefulness [7-10] in the practical application in the field of heat transfer. In this paper, from the fundamental principles of thermodynamics [11-13] and with the help of statistical mechanics, we reveal the improperness and incorrectness of such a concept.

The rest of the paper is organized as follow. In Sec. II, we prove the uniqueness of the integration factor $1/T$ and the resultant entropy as the state function. In Sec. III, we further discuss the errors in the definition of the entransy. The conclusions are given in Sec. IV

## II. Proof of the uniqueness of the integration factor $1/T$

In this section, we will prove the uniqueness of the integration factor associated with the exchanged heat $\delta Q$. As an illustration, we first demonstrate the result in the ideal gas with the internal energy $U$, the temperature $T$, the volume $V$, and the pressure $P$.

According to the first law of thermodynamics, the changed heat of the gas reads $\delta Q = dU - \delta W$, which can be further written as (with the work done on the ideal gas $\delta W = -PdV$)

$$\delta Q = dU + PdV. \tag{1}$$

An infinitesimal change of a generic thermodynamic function $\Lambda$ can be defined as $\delta \Lambda = f(T)\delta Q$, namely,

$$\delta \Lambda = f(T)(dU + PdV). \tag{2}$$

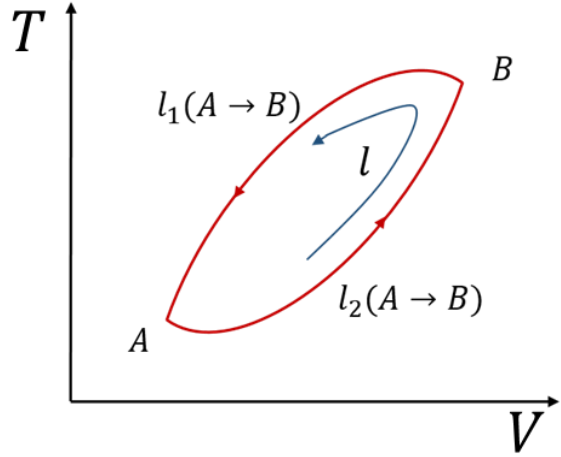

**Figure 1**. The evolution path of the ideal gas in the parameter space $T - V$

As illustrated in Fig. 1, the state $(T_0, V_0)$ at point A can be connected by two paths ($l_1$ and $l_2$) to the state $(T_f, V_f)$ at point B in the $T - V$ space. If $\Lambda$ is a state function, it is required that $\int_{A\to B}^{l_1} \delta\Lambda = \int_{A\to B}^{l_2} \delta\Lambda$ [2]. Accordingly, the loop integral of $\delta\Lambda$ in the $T - V$ space is strictly zero, namely,

$$\oint \delta\Lambda = \int_{A\to B}^{l_1} \delta\Lambda + \int_{B\to A}^{l_2} \delta\Lambda = \int_{A\to B}^{l_1} \delta\Lambda - \int_{A\to B}^{l_2} \delta\Lambda = 0. \tag{3}$$

With Eq. (2), the above equation can be specifically written as

$$\Delta\Lambda = \oint \delta\Lambda = \oint f(T)\left(C_V dT + \frac{nRT}{V}dV\right) = 0, \tag{4}$$

where we have used $dU = C_V dT$ and the equation of state of the ideal gas $PV = nRT$. Here, $C_V$ is the heat capacity at a constant volume, $n$ is the number of moles of the gas, and $R$ is the gas constant. With Green's theorem, the loop integral in Eq. (3) can be rewritten in the form of the surface integration as

$$\Delta \Lambda = \iint \left\{ \frac{\partial}{\partial V}[f(T)C_V] - \frac{\partial}{\partial T}\left[f(T)\frac{nRT}{V}\right]\right\} dV dT = 0. \tag{5}$$

This equation is valid irrespective of the integral intervals in the $T - V$ space. Hence, we have

$$f(T)\frac{\partial C_V}{\partial V} - \frac{\partial}{\partial T}\left[f(T)\frac{nRT}{V}\right] = 0. \tag{6}$$

The fact $\partial C_V / \partial V = 0$ for the ideal gas further simplifies the above equations to

$$\frac{\partial}{\partial T}\left[f(T)\frac{nRT}{V}\right] = 0, \tag{7}$$

whose solution can be uniquely determined as

$$f(T) = \frac{\alpha}{T}, \tag{8}$$

with $\alpha$ being an arbitrary constant independent of the temperature $T$. Without losing generality, we choose $\alpha = 1$, i.e., $\delta\Lambda = \delta Q/T$. Thus, $\delta\Lambda$ is nothing but the change of the thermodynamic entropy $dS$. In summary, we prove that for the classical ideal gas only one state function associated with the exchanged heat, namely the entropy, can be defined. A similar proof for such an ideal gas system was proposed by Weiss [3].

Having shown the uniqueness of the integration factor associated with the exchanged heat for the classical ideal gas, in the following, we will prove a theorem on the uniqueness of the integration factor associated with $\delta Q$ for a generic thermodynamic system.

***Theorem:*** For a generic thermodynamic system, the universal thermodynamic state function $\Lambda$ can be defined as $d\Lambda = f(T, \lambda)\delta Q$, if and only if $f(T, \lambda) = \alpha/T$ with α, a system-independent constant.

***Proof:*** For a generic thermodynamic system with the internal energy $U$, the first law of thermodynamic law reads

$$\delta Q = dU - \delta W = dU - Y d\lambda, \tag{9}$$

where $Y$ and $\lambda$ are the generalized force and the generalized displacement, respectively. Similar to the discussions for the ideal gas system, the condition for $\delta\Lambda = f(T,\lambda)\delta Q$ to be an exact differential ($\Lambda$ to be a state function) is: For an arbitrary loop in the $T - V$ space, we always have

$$\Delta\Lambda = \oint f(T,\lambda)\delta Q = \oint f(T,\lambda)(dU - Yd\lambda) = 0. \tag{10}$$

Noticing the relation for the internal energy

$$dU = \frac{\partial U}{\partial T}dT + \frac{\partial U}{\partial \lambda}d\lambda, \tag{11}$$

Eq. (10) can be rewritten in the form of the surface integration with Green's theorem as

$$\Delta\Lambda = \iint \left\{\frac{\partial}{\partial \lambda}\left[f(T,\lambda)\frac{\partial U}{\partial T}\right] - \frac{\partial}{\partial T}\left[f(T,\lambda)\left(\frac{\partial U}{\partial \lambda} - Y\right)\right]\right\} d\lambda dT = 0. \tag{12}$$

The above equation is valid irrespective of the integration intervals. Hence, we have

$$\frac{\partial}{\partial \lambda}\left[f(T,\lambda)\frac{\partial U}{\partial T}\right] - \frac{\partial}{\partial T}\left[f(T,\lambda)\left(\frac{\partial U}{\partial \lambda} - Y\right)\right] = 0, \tag{13}$$

which can be further simplified as

$$\frac{\partial U}{\partial T}\frac{\partial f(T,\lambda)}{\partial \lambda} + f(T,\lambda)\frac{\partial Y}{\partial T} + \left(Y - \frac{\partial U}{\partial \lambda}\right)\frac{\partial f(T,\lambda)}{\partial T} = 0. \tag{14}$$

In statistical mechanics, the internal energy and the generalized force can be written as [1]

$$U = \sum_{j=1}^{j=N} p_j E_j, \tag{15}$$

and

$$Y = \left(\sum_{j=1}^{j=N} p_j \frac{dE_j}{d\lambda}\right), \tag{16}$$

where $E_j = E_j(\lambda)$ ($j = 1,2\cdots N$) is the $j$-th energy level of the system [*For simplicity, we consider a system with discrete energy levels. But it is straightforward to extend our discussions to systems with a continuous energy spectrum.*], and it is $\lambda$-dependent.

$$p_j = \frac{e^{-\beta E_j(\lambda)}}{\sum_{j=1}^{j=N} e^{-\beta E_j(\lambda)}} \equiv \frac{e^{-\beta E_j}}{Z} \tag{17}$$

is the corresponding thermal equilibrium distribution of the system on the $j$-th energy level with $\beta = 1/(k_B T)$ as the inverse temperature and $k_B$ as the Boltzmann constant. Combining Eqs. (15-17), we find

$$\frac{\partial U}{\partial T} = \frac{\partial}{\partial T}\left(\sum_{j=1}^{j=N} \frac{e^{-\beta E_j}}{Z} E_j\right) = \frac{\langle E_j^2 \rangle - U^2}{k_B T^2} \equiv C_\lambda, \tag{18}$$

$$\frac{\partial Y}{\partial T} = \frac{\partial}{\partial T}\left(\sum_{j=1}^{j=N} \frac{e^{-\beta E_j}}{Z} \frac{dE_j}{d\lambda}\right) = \frac{\beta}{T}\left(\sum_{j=1}^{j=N} p_j E_j \frac{dE_j}{d\lambda} - UY\right), \tag{19}$$

and

$$\frac{\partial U}{\partial \lambda} = \frac{\partial}{\partial \lambda}\left(\sum_{j=1}^{j=N} \frac{e^{-\beta E_j}}{Z} E_j\right) = Y + \beta UY - \beta \sum_{j=1}^{j=N} p_j E_j \frac{dE_j}{d\lambda}. \tag{20}$$

Substituting the above three relations into Eq. (14), we obtain

$$C_\lambda \frac{\partial f(T,\lambda)}{\partial \lambda} + \Theta\left(\frac{\partial f(T,\lambda)}{\partial T} + \frac{f(T,\lambda)}{T}\right) = 0, \tag{21}$$

where

$$\Theta \equiv \beta\left(\sum_{j=1}^{j=N} p_j E_j \frac{dE_j}{d\lambda} - UY\right) = T\frac{\partial Y}{\partial T} \tag{22}$$

is determined by the equation of state $Y = Y(T,\lambda)$ of the system. With the assumption of the factorized structure $f(T,\lambda) = g(T)h(\lambda)$, Eq. (21) becomes

$$C_\lambda \frac{d\ln h(\lambda)}{d\lambda} + \Theta\left(\frac{d\ln g(T)}{dT} + \frac{1}{T}\right) = 0. \tag{23}$$

The solution to Eq. (23) follows:

$$\left(\frac{d\ln g(T)}{dT} + \frac{1}{T}\right) = \mu = -\frac{C_\lambda}{\Theta}\frac{d\ln h(\lambda)}{d\lambda} \tag{24}$$

where $\mu = \mu(T)$ is a function of $T$ independent of $\lambda$. The solution to Eq. (24) is

$$h(\lambda) = h_0 e^{-\mu \int \frac{\Theta}{C_\lambda} d\lambda}, g(T) = \frac{g_0}{T} e^{\int \mu dT}. \tag{25}$$

And the integration factor $f(T, \lambda)$ can be explicitly written as

$$f(T, \lambda) = \frac{\alpha}{T} e^{-\mu \int \frac{\Theta}{C_\lambda} d\lambda + \int \mu dT}. \tag{26}$$

Here, $g_0, h_0$ are integral constants independent of $T$ and $\lambda$, and $\alpha = g_0 h_0$. In order to be consistent with the factorized structure assumption for $f(T, \lambda)$, $\Theta/C_\lambda$ also needs to have a factorized structure of $T$ and $\lambda$. In the equation above, the form of $f(T, \lambda)$ is not unique due to the multiple choice of the function $\mu(T)$. The dependence of $f(T, \lambda)$ on $\Theta/C_\lambda$ with the specific thermodynamic heat capacity and the equation of state prohibit the definition of the universal quantity. Therefore, $\mu$ can only be set to be $\mu = 0$ to make $\Lambda$ a universal state function

$$f(T, \lambda) = \frac{\alpha}{T}. \tag{27}$$

We have proven the uniqueness theorem of the integration factor associated with the exchanged heat. Without loss of generality we set $\alpha = 1$, and we obtain

$$\delta \Lambda = \frac{\delta Q}{T} = dS, \tag{28}$$

which indicates that, associated with the exchanged heat $\delta Q$, the entropy defined via $dS = \delta Q/T$ is the only universal state function. The existence of the integration factor $1/T$ for the exchanged heat is known in thermodynamics. We have shown no other integration factors exist for the exchanged heat $\delta Q$.

The entransy $\delta E_{vh} = T \delta Q$ is introduced as a "state function" for the purpose of optimizing the heat transfer [5]. The supporters of the entransy claim it as a new state function by regarding a new integration factor $T$, which contradicts the theorem. The theorem directly excludes the entransy as a state function. Clearly, the entransy is essentially different from the well-defined thermodynamic state quantities such as the internal energy, the free energy, and the entropy. Thus, we conclude that the entransy introduced in Ref. [5] cannot be regarded as a fundamental thermodynamic quantity. And the entransy is introduced for the system with a fixed volume [5]. Such assumption leaves the temperature $T$ as the only variable. It is meaningless to talk about the state function of a single-valued function since in thermodynamics a state function is exclusively a function of two or more variables.

For the case of the single-valued function, the entransy is written explicitly by the integration $E_{vh} = \int_0^T C_V(T)TdT$. Alternatively, the definition of the entransy [5] via the internal energy is given as $E_{vh} = UT/2 = C_V T^2/2$, with the analogy to the definition of the energy of the electronic capacity. To assure the equivalence of the two definitions above, the capacity at a constant volume is assumed as a constant to simplify the integration $E_{vh} = \int_0^T C_V(T)TdT = C_V T^2/2$. However, such assumption is only valid for some systems, such as classical ideal gas, in the high temperature regime, but not valid for any real solid-state materials in an arbitrary circumstance. One typical heat capacity $C_V$ as the function of temperature $T$ is known as the Debye's law, which shows that, in the low temperature regime of $T \ll \Theta_D$, $C_V \propto T^3$ rather than a constant [14]. Here $\Theta_D$ is the Debye temperature. Such oversimplified assumption prevents the practical application.

## III. Conclusions

In summary, it is a common sense that $1/T$ is an integration factor associated with the exchanged heat $\delta Q$. The uniqueness of this integration factor has been proven for some specific systems with given equations of state [3]. In this paper, from the perspective of statistical mechanics, we prove the uniqueness of $1/T$ as the integration factor associated with the exchanged heat $\delta Q$ for a generic thermodynamic system without referring to its equation of state. Such a theorem prevents the possibility of defining any new state function associated with the exchanged heat other than the entropy. With this theorem, we clearly exclude the possibility of the entransy as a state function in thermodynamics. In addition, we have also shown errors in the derivation of the entransy with the false assumption of a temperature-independent heat capacity. We conclude that the entransy cannot be a state function in thermodynamics and the false assumption in the derivation prevents its practical applications in heat transfer in any real materials.

## Acknowledgments

Y. H. Ma is grateful to R. X. Zhai and Z. Y. Fei for helpful discussions. This work was supported by the National Natural Science Foundation of China (NSFC) (Grants No. 11534002, No. 12088101, No. 11875049, No. U1730449, No. U1930402, No. U1930403, No. 11775001, No. 11534002, and No. 11825001), and the National Basic Research Program of China (Grants No. 2016YFA0301201).

## Declaration of Competing Interest

The authors declared that they have no conflicts of interest to this work.